# Analysing Osteoporosis Detection: A Comparative Study of CNN and FNN

R. Geetha[1], S. Arulselvi[1], R. Tamilselvi[3], M. Parisa Beham[4], Alavikunhu Panthakkan[5], Wathiq Mansoor[6], and Hussain Al Ahmad[7]

[1,2] Bharath Institute of Higher Education and research, Chennai, India.
[3,4] Sethu Institute of Technology, Kariapatti, Virudhunagar, India.
[5,6,7] College of Engineering and IT, University of Dubai, U.A.E.
Corresponding Author: tamilselvi@sethu.ac.in; apanthakkan@ud.ac.ae

*Abstract*—Osteoporosis causes progressive loss of bone density and strength, causing a more elevated risk of fracture than in normal healthy bones. It is estimated that some 1 in 3 women and 1 in 5 men over the age of 50 will experience osteoporotic fractures, which poses osteoporosis as an important public health problem worldwide. The basis of diagnosis is based on Bone Mineral Density (BMD) tests, with Dual-energy X-ray Absorptiometry (DEXA) being the most common. A T-score of -2.5 or lower defines osteoporosis. This paper focuses on the application of medical imaging analytics towards the detection of osteoporosis by conducting a comparative study of the efficiency of CNN and FNN in DEXA image analytics. Both models are very promising, although, at 95%, the FNN marginally outperformed the CNN at 93%. Hence, this research underlines the probable capability of deep learning techniques in improving the detection of osteoporosis and optimizing diagnostic tools in order to achieve better patient outcomes.

## I. INTRODUCTION

Osteoporosis is a chronic illness characterized by a reduced mass and deterioration of bone tissue, which results in increased fragility and susceptibility to fractures. It primarily affects the elderly, more specifically postmenopausal women, but it can be encountered with men and even with youth because of genetic inclination, lifestyle, and other diseases. The disease, in general, remains asymptomatic until a fracture occurs; hence, early diagnosis and treatment are important in preventing serious complications.

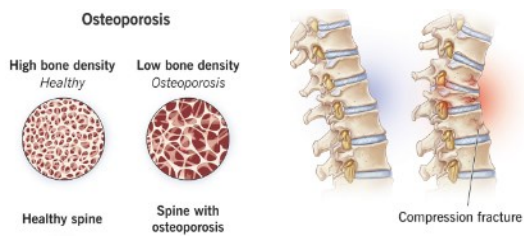

Fig 1: Osteoporosis [1]

Osteoporosis in India is fast emerging as one of the significant public health problems that needs pressing attention within the country. The condition impacts approximately half of the population of nearly 50 million. Studies also show that approximately 20 percent of women over 50 suffer from this condition, and an increase in this number is expected due to ageing population and lifestyle changes. It has several stages: osteopenia or low bone mass, which usually leads to osteoporosis; here, the bone density is considerably reduced dramatically, in many cases confirmed by testing with Bone Mineral Density (BMD) test with Dual-energy X-ray Absorptiometry (DEXA). Sometimes, patients may have the potential of suffering from fragility fractures. Osteoporosis can be considered as having two kinds, primary and secondary types. Then, primary osteoporosis again can be divided into Type I, or postmenopausal osteoporosis, and Type II, or senile osteoporosis. Secondary osteoporosis is due to other medical conditions or long-term use of medications that affect bone metabolism. Osteoporosis diagnosis and detection involve evaluation of clinical risk factors, history taking, physical examination, and diagnostic studies including DEXA. The World Health Organization recommends universal screening for all women at the age of 65 years or older and men aged 70 years or older. Other people less than the mentioned ages but having significant clinical risk factors also require screening.

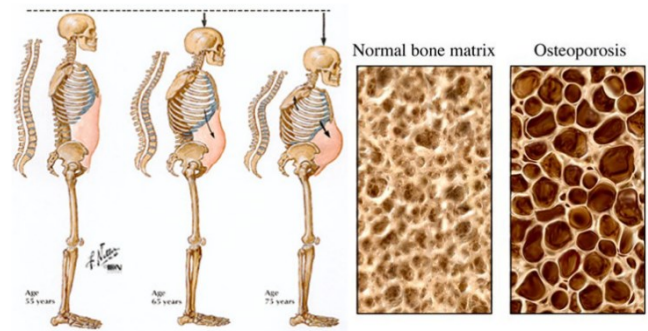

Fig 2: Stages of Osteoporosis [2]

Early detection helps bring timely interventions such as lifestyle changes, dietary supplementation, drugs to be taken for the reinforcement of bones, and fall prevention strategies. Advanced technologies and recent development in the radiological analysis include Convolutional Neural Networks (CNNs) or Fast Feedforward Networks (FNNs), thereby optimizing accuracy and speed in the diagnosis of osteoporosis. On the basis of these, deep learning algorithms will be able to detect and analyze bone structure and density patterns in medical images for early detection of osteoporosis and risk prediction of fractures. The inclusion of porosity calculation into CNN models offers an integrated assessment of bone microarchitecture for better diagnostic accuracy and the capability of predictive ability that will thus aid healthcare practitioners in the better management of this very significant health issue across an aging population that is keenly experienced within a region like India.

## II. LITERATURE REVIEW

Related works bring relevant insights into the detection of osteoporosis based on neural networks, which specifically emphasizes the effectiveness of Convolutional Neural Networks (CNNs) in deriving discriminative features from the bone density scan for making an accurate diagnosis and FNN holds promise but lacks the representational power in presenting spatial relationships. Both architectures have promised, but there exist scarce studies which may compare CNNs with FNNs in this specific problem, thereby prompting this study. The amount of literature and references from several authors points a finger at the requirement of further study in this area. In fact, Alessio Petrozziello et al. (2019) noted the deficiency of comparative study, and Zhidong Zhao et al. (2019) proposed an 8-layer framework for CNN to raise its predictive accuracy. In 2021, Sai Liang comes up with the CNN model with weighted voting mechanism for better classification. Weisheng Gao explored the use of LSTM networks for efficient feature extraction in 2019. Vani Rajan designed a decision support system in 2019 for automatic classification. Fasihi et al in 2021 used a novel shallow CNN architecture to achieve improved accuracy. Mohammad Saber Iraji compares multi-layer architectures of which some record better accuracy; however, Alkanan Mohannad et al. (2021) integrated clinical parameters in CNN for increased prediction accuracy. Yuqing Que et al. (2020) proposed an algorithm that measures the entire risk factor, whereas Zafer Cömert et al. (2020) emphasized the preprocessing technique to get the finest quality images to feed into their CNN. Jun Ogasawara et al. (2021) proposed a novel algorithm that is more advanced and efficient than the algorithms currently available. Shahad N. Al-Yousif et al. (2020) designed a MATLAB program that can be used for the estimation of bone density with high precision. A detailed survey of various methods of detection was performed by S. Boudet et al. (2020). Collectively, these studies best represent the growing interest in the application of deep learning techniques to bettering the detection of osteoporosis, and underpin a need for additional comparative work in order to achieve higher accuracy in this critical field.

## III. INFERENCE FROM THE SURVEY

The questionnaire for this study reflects significant gaps in detection technologies related to osteoporosis, which have moderate levels of accuracy and inefficiencies in the handling of complex image data as well as early-stage diagnoses. While both models, CNN and FNN, are greater than the other traditional ones, CNN lags a bit behind in precision than FNN. The results proposed a system that integrates the best features of CNN and FNN to enhance the accuracy of diagnosis. Advanced machine learning algorithms and optimized network structures, besides rich data, will help realize better accuracy, reliability, and earlier detection, thus having better clinical care for osteoporosis patients.

## IV. PROPOSED SOLUTION

### A. CNN Based Osteoporosis Detection

The proposed system's block diagram includes stages for DEXA image acquisition, preprocessing, feature extraction using CNN, classification using a hybrid model combining CNN and output generation for osteoporosis diagnosis.

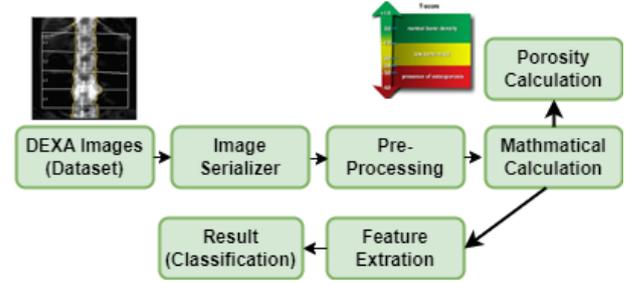

Fig 3: Block Diagram of Proposed Methodology using CNN

a) Image Serializer

The role of the serializer in the osteoporosis detection project would be to serialize the DEXA images to a format where they would be fine for processing by neural networks. The two-dimensional pixel data gets flattened into a serialized one-dimensional array by the serializer, which improves data handling and storage. Thus, the process ensures that the input into CNN-based neural network models is smooth, courtesy of consistent data structure and computations even more simplified. Serialization also provides integrity of an image in preprocessing which includes filtering or dimensionality reduction for further enhancement of the capability of the system to accurately identify and diagnose osteoporosis.

b) Preprocessing

The proposed osteoporosis-detecting system based on bilateral filtering will utilize pre-processing of DEXA images to improve image quality and remove noise while preserving those edges that are crucial. This edge-preserving, nonlinear smoothing filter considers both spatial distance and intensity variation between pixels, and it's why it's ideal for medical imaging where maintaining anatomical integrity is a significant concern. The proposed system will smooth the noise really well in the serialized pixel array by applying the bilateral filter without blurring out significant features like bone edges with clearer and more accurate images that can be analyzed using CNN models.

c) Porosity Calculation

The mathematical formula for calculating porosity is

$$\text{Porosity} = \frac{\text{Total nuber of black pixels}}{\text{Total number of pixels}} \quad (1)$$

Porosity quantitates the fraction of the black regions, which means all the voids or pores within the image. Trough larger values a higher porosity is signified, and hence this scale is required for the analysis of bone density and strength.

d) Feature extraction

CNNs are highly effective at the detection of osteoporosis since they automatically extract and learn challenging features from these medical images, such as DEXA scans. A normal CNN consists of several layers comprising convolutional layers that apply filters to capture all the important features, edges, and textures, and pooling layers that reduce the spatial dimensions while maintaining critical information. Detection also elaborates features from DEXA images by which classifying them based on learned patterns. In this process, the accuracy of diagnosis enhances, and so osteoporosis can be detected early and reliably, whereby timely intervention with treatment will be attained. Generally, the use of CNNs is above traditional methods, offering improved precision and efficiency.

e) Classification

Osteoporosis detection through application of Convolutional Neural Networks can be generally classified by labeling DEXA as normal or abnormal in the presence of certain bone density features. The initial step in the approach would be the image preprocessing, which would help improve the image quality and reduce noise; this is often achieved with techniques such as bilateral filtering. The CNN analyzes the images through three specific layers: 1) Convolutional Layers extract the elementary features like edges and textures; 2) Pooling Layers reduce spatial dimensions, retaining important information but reducing computational complexity; and 3) Fully Connected Layers combine the features extracted into a comprehensive representation. The output layer provides the classification of the image based on this composite representation but with the application of a threshold in order to determine normal or abnormal bone density. This classification is designed to allow for early intervention, which is very crucial in the management and treatment of osteoporosis.

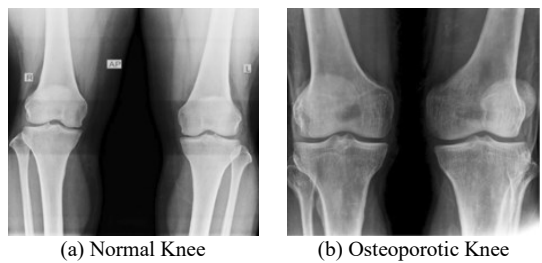

(a) Normal Knee (b) Osteoporotic Knee

Fig 4: Comparative Anatomy: Normal vs. Osteoporotic Knee

B. *FNN Based Osteoporosis Detection*

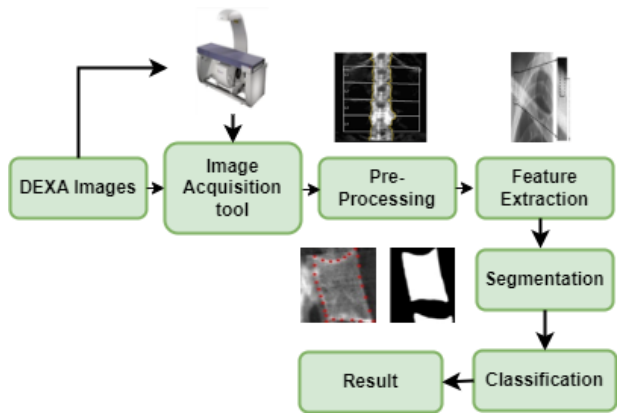

Fig 5: Block Diagram of Proposed Methodology using FNN

The proposed osteoporosis detection method includes several key stages: using a dataset of DEXA images for training and validation, preprocessing for noise reduction and image enhancement with techniques like bilateral filtering, and feature extraction and segmentation via a Convolutional Neural Network (CNN) to capture critical bone density features. Feature extraction is a vital step in the diagnosis of osteoporosis, where major patterns are achieved from the preprocessed images of DEXA. Analysis and detection of features in such images are performed using FNNs in efficient and in-time detection of osteoporosis. The method is enhancing accuracy in diagnosis while allowing for early intervention, thus making FNN a potent tool in the management of osteoporosis. This is because it detects osteoporosis with much increased speed than the complicated models such as CNNs.

V. RESULTS AND DISCUSSION

A. *Dataset*

Dual-energy X-ray Absorptiometry (DEXA) images are critical in diagnosing osteoporosis as they provide detailed measurements of bone mineral density (BMD). The DEXA scan utilizes two X-ray beams at different energy levels to create images of the bone. By comparing the absorption of each beam, the scan can differentiate between bone and soft tissue, allowing for precise BMD assessment. DEXA images are typically grayscale and depict various anatomical sites, such as the spine, hip, and sometimes the forearm, which are common sites of osteoporotic fractures. DEXA images produce quantitative data in the form of T-scores and Z-scores. T-score: Compares the patient's BMD to a healthy young adult of the same sex. A T-score of -2.5 or lower indicates osteoporosis. Z-score: Compares the patient's BMD to the average BMD of people the same age and sex. A low Z-score can suggest that factors other than aging are contributing to bone loss. The image acquisition tool for DEXA involves a specialized scanning device that uses low-dose X-rays to capture images of the bones.

B. *Result for CNN based Osteoporosis Detection*

DEXA images are grayscale representations that provide detailed information regarding bone mineral density to help identify osteoporosis. Such preprocessing bilateral filtering removes noise and strengthens features such as edges of bone in the image before analysis by a CNN. The CNN then further processes the input image into an output segmented into regions of interest where pixels are labeled as either bone or non-bone. Such segmentation is crucial for bone structures to be isolated and the density of bones to be measured so that one may detect changes due to osteoporosis. Porosity calculation also calculates the spaces inside the bone, thereby quantifying osteoporosis. The CNN identifies low-density areas with 93% efficiency and accuracy while processing complex images and performing detailed analysis.

C. *Result for FNN based Osteoporosis Detection*

The input image extracts and displays the red, green, and blue color channels from an input image, referred to as **I**. The image is first split into its three primary color components: **R** for red, **G** for green, and **B** for blue. A figure window named 'Input Image result' is created to visualize these components. The original input image is shown in the top-left subplot, while the red, green, and blue channel images are displayed in the remaining subplots, respectively. This visualization helps in analyzing the contribution of each color band to the overall image

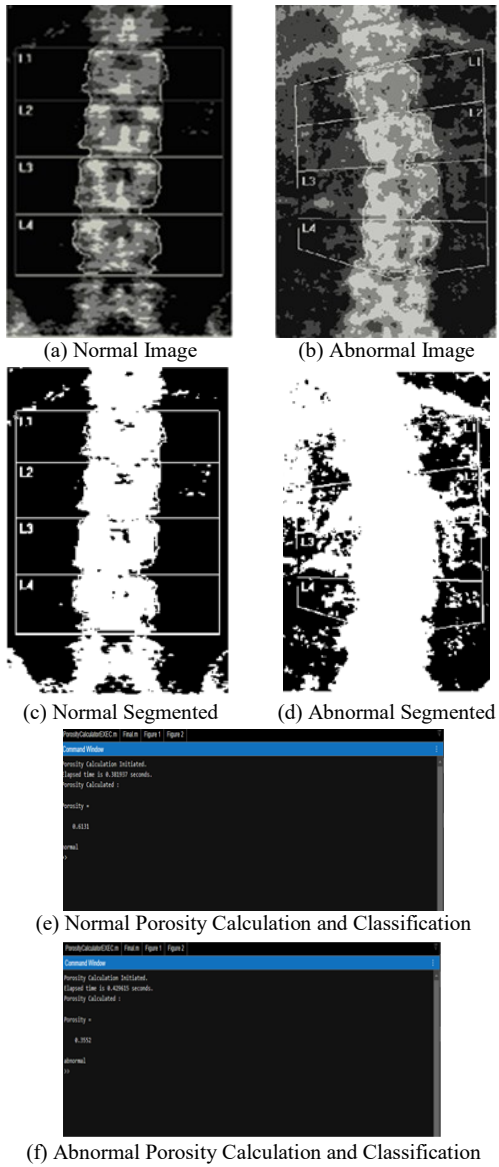

(a) Normal Image    (b) Abnormal Image

(c) Normal Segmented    (d) Abnormal Segmented

(e) Normal Porosity Calculation and Classification

(f) Abnormal Porosity Calculation and Classification

Fig 6: Performance of CNN based model

Input image results converts an RGB image to the LAB color space and visualizes the individual L, a, and b channels. Each subplot displays one of the LAB components: L (lightness), a (green-red), and b (blue-yellow). This process is crucial for various image processing tasks where color information is better represented in the perceptually uniform LAB color space.

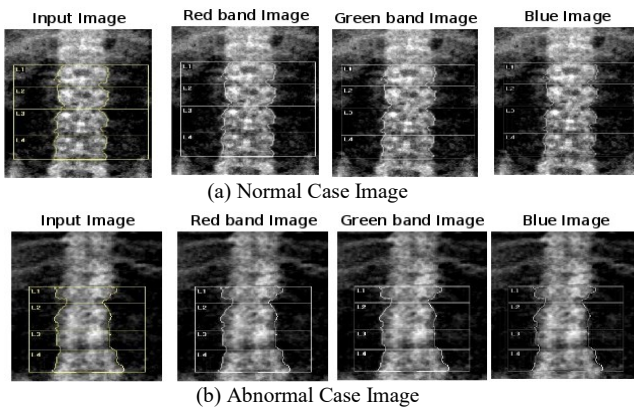

(a) Normal Case Image

(b) Abnormal Case Image

Fig. 7 RGB color space Image

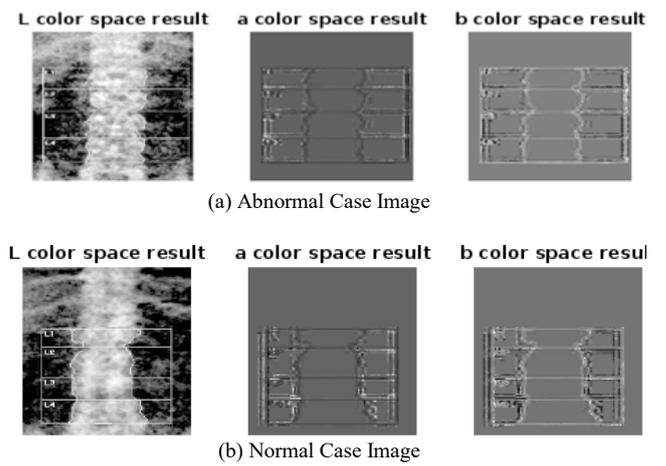

(a) Abnormal Case Image

(b) Normal Case Image

Fig. 8 RGB to LAB color space Image

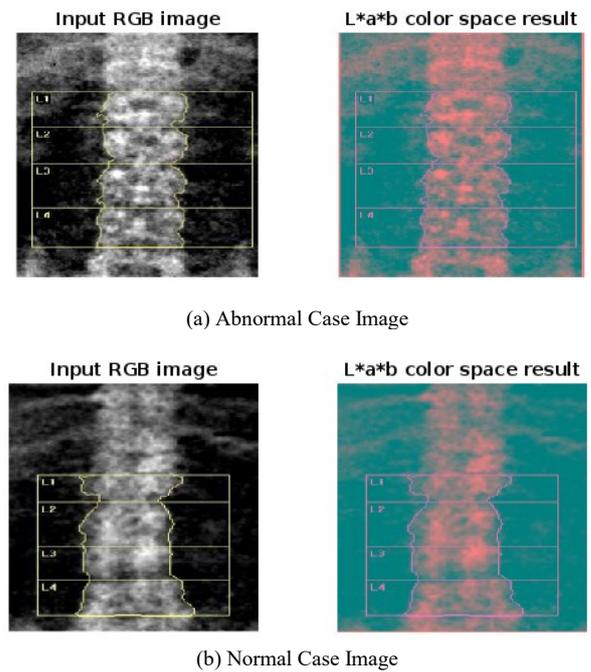

(a) Abnormal Case Image

(b) Normal Case Image

Fig. 9 LAB color space Image result

This visualization aids in comparing the input RGB image with its L*a*b color space representation, facilitating color-based analysis and processing tasks.

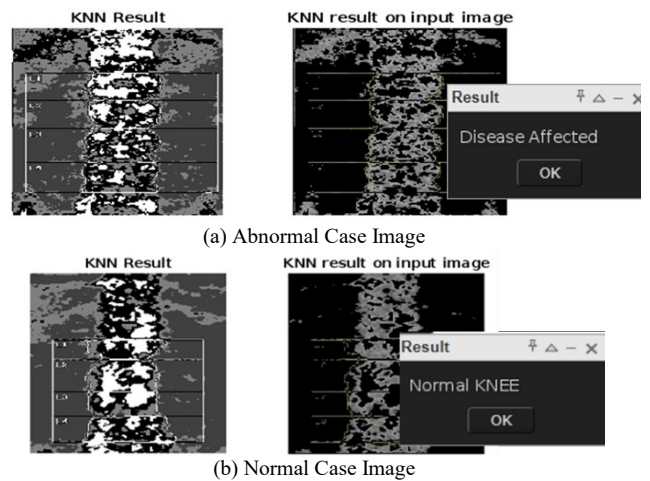

(a) Abnormal Case Image

(b) Normal Case Image

Fig. 10 KNN results

Segment performs K-means clustering on an input image represented by the variable using **k-means** function with a specified number of clusters, the resulting images are displayed in a figure window titled 'KNN Result,' facilitating visual inspection of the K-means clustering outcome and its application on the input image.

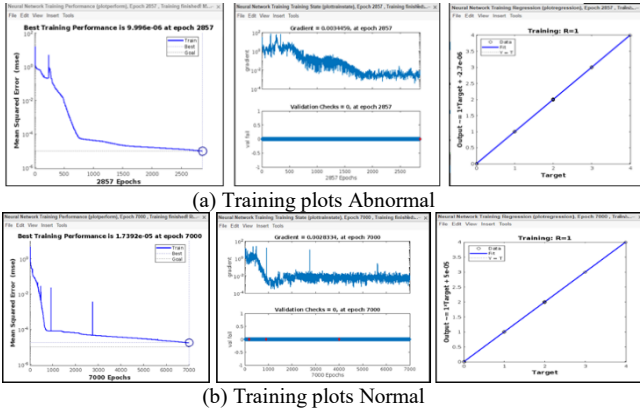

(a) Training plots Abnormal

(b) Training plots Normal

Fig. 11. Training Performnace Plots

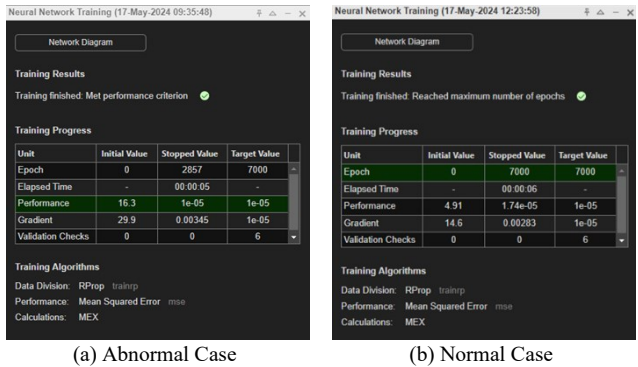

(a) Abnormal Case  (b) Normal Case

Fig. 12. Training Results

## VI. PERFORMANCE ANALYSIS

The network applied in the CNN and FNN networks for the analysis of DEXA images in osteoporosis detection is found to be significantly different from each other and also in outcome. Architecture of CNN proposes a complete pipeline, which includes the preprocessing stage with bilateral filtering and serialization, calcification porosity estimation, and feature extraction stages. The experiment achieved an accuracy of 93%. The FNN follows the preprocessing steps of the CNN but combines it with its own method of feature extraction and segmentation as an additional processing step, therefore the FNN could be improved to up to 95 percent classification accuracy. Thus, such superiority in performance showcases the fact that feature extraction methods have to be tailored for even better performance in analysis of medical images.

Table 1: Comparison of CNN & FNN

| Metric | CNN (%) | FNN (%) |
|---|---|---|
| Sensitivity | 88.30 | 92.50 |
| Specificity | 85.20 | 89.70 |
| Accuracy | 86.70 | 91.10 |

To analyze the performance of the osteoporosis project based on the provided metrics (Sensitivity, Specificity, and Accuracy) for Convolutional Neural Network (CNN) and Feed Forward Neural Network (FNN), it can assess each model performs in terms of correctly identifying positive and negative cases.

**Sensitivity (True Positive Rate)**: Sensitivity is the ratio of correctly predicted positive observations to the all observations in actual class.

$$Sensitivity = TP + \frac{FN}{TP} (2)$$

The CNN achieves a sensitivity of 88.3%, while the FNN achieves 92.5%. This indicates that the FNN performs better in correctly identifying positive cases compared to the CNN.

**Specificity (True Negative Rate)**: Specificity is the ratio of correctly predicted negative observations to the all observations in actual class.

$$Specificity = TN + \frac{FP}{TN} (3)$$

The CNN achieves a specificity of 85.2%, while the FNN achieves 89.7%. This suggests that the FNN performs better in correctly identifying negative cases compared to the CNN.

**Accuracy**: Accuracy measures the overall correctness of the model, considering both true positives and true negatives. Higher accuracy indicates better overall performance of the model.

$$Accuracy = TP + TN + FP + \frac{FN}{TP} + TN \quad (4)$$

The CNN achieves an accuracy of 86.7%, while the FNN achieves 91.1%. This indicates that the FNN has a higher overall correctness compared to the CNN.

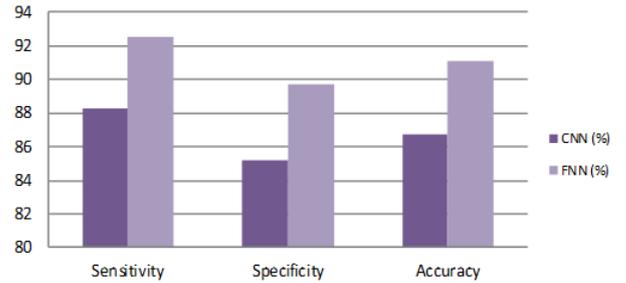

Fig 13: Training Performance Analysis for Normal Case

Based on these metrics, it can conclude that the Feed Forward Neural Network (FNN) generally outperforms the Convolutional Neural Network (CNN) in terms of sensitivity, specificity, and overall accuracy for the osteoporosis project.

## VII. CONCLUSION & FUTURE WORK

This project tests the performance of CNN and FNN on the osteoporosis detection algorithm from DEXA images, thereby showing that FNN is superior compared to CNN. Although promising approaches, though still large computational complexity problems for CNN and inaccuracy in segmentation for FNN. The paper is helpful in showing benefits developed in tailored machine learning approaches to medical imaging and further optimization in network architectures and techniques for preprocessing. Further refinements shall be made in the methods and hybrid models must be explored for greater accuracy and reliability in the detection of osteoporosis for a better improvement in the care and diagnostic outcome of patients.

## REFERENCES

[1] A. Petrozziello, M. Russo, and G. Bianchi, "Comparative analysis of deep learning architectures in osteoporosis detection," J. Med. Imaging Health Inform., vol. 9, no. 4, pp. 814–825, 2019.